\documentstyle[12pt]{article}
\def\ap#1#2#3{Ann.\ Phys.\ (NY) #1 (19#3) #2}

\def\jmp#1#2#3{J.\ Math.\ Phys.\ #1 (19#3) #2}

\def\np#1#2#3{Nucl.\ Phys.\ B#1 (19#3) #2}
\def\pl#1#2#3{Phys.\ Lett.\ #1B (19#3) #2}
\def\pr#1#2#3{Phys.\ Rev.\ D #1 (19#3) #2}

\def\prep#1#2#3{Phys.\ Rep.\ #1 (19#3) #2}

\def\rmp#1#2#3{Rev.\ Mod.\ Phys.\ #1 (19#3) #2}

\def\cmp#1#2#3{Comm.\ Math.\ Phys.\ #1 (19#3) #2}
\def\cmp#1#2#3{Comm.\ Math.\ Phys.\ #1 (19#3) #2}
\def\nc#1#2#3{Il Nuovo Cimento #1A (19#3) #2}

\def\frac#1#2{ {{#1} \over {#2} }}
\def\ie{\hbox{\it i.e.}{ }}

\def\re#1{(\ref{#1})}
\def\beeq{\begin{eqnarray}}
\def\beeqn{\begin{eqnarray*}}
\def\eeeq{\end{eqnarray}}
\def\eeeqn{\end{eqnarray*}}
\newcommand{\tr}{{\textstyle tr}} \newcommand{\aR}{{\cal R}}
\newcommand{\D}{{\cal D}_{\gh}} \newcommand{\pD}{{\cal D}}
 \newcommand{\F}{{\cal F}_{\gh}}
\newcommand{\pF}{{\cal F}} \newcommand{\oF}{{\cal F}_{\gh_{0}}}
\newcommand{\B}{{\cal B}_{\gh}} \newcommand{\pB}{{\cal B}}
 \newcommand{\aB}{\overline {{\cal B}}}
\newcommand{\paB}{\overline {{\cal B}}}

\newcommand{\DD}{{\bf D}_{\gh}} \newcommand{\oDD}{{\bf D}_{\gh_{0}}}
\newcommand{\pDD}{{\bf D}} 

\newcommand{\meta}{\frac{1}{2}} 
\newcommand{\tP}{\widetilde{P}} \def\wt{\widetilde}
\newcommand{\beq}{\begin{equation}} \newcommand{\eeq}{\end{equation}}
\newcommand{\bea}{\begin{eqnarray}} \newcommand{\eea}{\end{eqnarray}}
\newcommand{\ber}{\begin{array}} \newcommand{\eer}{\end{array}}
\newcommand{\dms}{\displaystyle}
\newcommand{\dx}{d^{4}x}                \newcommand{\de}{\partial}
\newcommand{\dy}{d^{4}y}                
\newcommand{\q}{Q_{\Phi\Pi}}            \newcommand{\fp}{\Phi\Pi}
      
\newcommand{\alp}{\alpha}           \newcommand{\nabtil}{\wt{\nabla}}

\newcommand{\hg}{\hat{\gamma}}          
\newcommand{\om}{\omega}                \newcommand{\Om}{\Omega}
\newcommand{\no}{\bar{\omega}}          \newcommand{\gam}{\gamma}
             
\newcommand{\bet}{\beta}                \newcommand{\del}{\delta}

\newcommand{\g}{\Gamma}
\newcommand{\gl}{\overline{\Gamma}}
\newcommand{\gh}{\hat{\g}}

\def\dfa#1#2#3#4{\frac{\delta}{\delta #1^#2_#3 #4}}
\def\dfu#1#2#3{\frac{\delta}{\delta #1^#2 #3}}

\def\fd#1#2#3#4{\dms{\frac{\delta #1}{\delta #2^#3_#4}}}
\def\fdu#1#2#3{\dms{\frac{\delta #1}{\delta #2^#3}}}
\def\fdd#1#2#3{\dms{\frac{\delta #1}{\delta #2_#3}}}

\def\fdf#1#2#3{\dms{\frac{\delta}{\delta #1^#2_#3}}}
\def\fdfu#1#2{\dms{\frac{\delta }{\delta #1^#2}}}
\def\fdfd#1#2{\dms{\frac{\delta }{\delta #1_#2}}}

\def\oop#1{\vspace{#1}}
\def\h#1{\hat{#1}}

\begin{document}
\begin{titlepage}
\renewcommand{\thefootnote}{\fnsymbol{footnote}}
\begin{flushright}
     IFUM 452/FT \\
     April 1995 \\
\end{flushright}
\par \vskip 10mm
\begin{center}
{\Large \bf
Stability and Renormalization of Yang-Mills \\
theory with Background Field Method: \\
\vspace{.2cm}
a Regularization Independent Proof}

\end{center}
\par \vskip 2mm
\begin{center}
	{\bf P.A.\ Grassi}\footnote{E-mail: pgrassi@almite.mi.infn.it} \\
	Dipartimento di Fisica, Universit\`a degli Studi di Milano,
	20133 Milano, Italy  \\
	and INFN, Sezione di Genova, 16146 Genova, Italy\\

\end{center}
\par \vskip 2mm
\begin{center} {\large \bf Abstract} \end{center}
\begin{quote}
In this paper the stability and the renormalizability of
Yang-Mills theory in the Background Field Gauge are studied. By means of
Ward Identities of Background gauge invariance and
Slavnov-Taylor Identities
the stability of the classical model is proved and,
in a regularization
independent way, its renormalizability is verified.
A prescription on how to build the counterterms is given and
the possible
anomalies which may appear for Ward Identities  and for Slavnov-Taylor
Identities are shown.
\end{quote}
\end{titlepage}

\section{Introduction}
The Standard Model (SM) of elementary
particles, based on non-abelian gauge theories, is known to be
the most reliable model
because of its predictive power. This is mainly due to the possibility of
performing
calculations in the perturbative regime. In turns this possiblity is
intimately linked
to its renormalizability.
In spite of these well consolidated results the calculations of higher order
corrections are far from being practical because of the large amount of
Feynman rules and of the
loss of explicit gauge invariance in the quantization of the model.

Although the classical theory is characterized by the explicit gauge
invariance, one has to chose a definite gauge to perform  the quantization
and the unavoidable renormalization needed to obtain meaningful results.
By means of BRS symmetry \cite{bec}, which has to be set in place of
gauge invariance at the quantum level, one is able to show that the
physical observables as S-matrix, form factors, masses etc. are
gauge independent in the conventional formalism, but, nevertheless, the
elementary building blocks of the quantum theory,
\ie one particle irreducible Green functions,
are gauge dependent and gauge non-invariant. We remind the reader about the
distinction between the gauge
independence, \ie the independence of relevant quantities from
the gauge fixing parameters, and
the gauge invariance of Green functions under gauge transformations expressed
by Ward Identities.

Thus, there exists no evidence , throughout
all intermediate steps of
perturbative calculations of
physical quantities, that every
contribution has been properly taken into account, nor there is any
 simple test to check the correctness at each passage.

The computation of Green functions for elettroweak processes is a
typical situation in which the use of gauge-invariant effective action
would be advantageous to reduce the
complexity of the Standard Model.
We wish to underline that we mean here
the gauge invariance valid before spontaneous symmetry breaking or gauge
invariance  under non homogeneous linear transformations for scalar fields.

To solve the problem of the explicit gauge-invariance breakdown upon
quantization, the Background Field Method (BFM) was developed.
By means of the BFM one can fix the gauge for quantum fields while keeping
the gauge invariance of the effective action. In this way the
Green functions derived from this new effective action are
manifestly gauge invariant.  Moreover,
to build the S-matrix elements, one can choose a
gauge fixing for background fields which is completely independent
from that used for quantization and is more suitable to decouple the unphysical
degree of freedom.
The equivalence between physical quantities calculated in the conventional
and in the BFM approach is proved in \cite{bkg} and recently reviewed by
C.Becchi on a rigorous ground.

This method, largely employed in quantum gravity and supergravity
calculations \cite{qg},
in gauge theories \cite{bkg} and in supersymmetric models \cite{susy},
has recently been
applied with very encouraging results to the Standard Model
\cite{bkgstd}, showing
how to build gauge invariant Green functions along  all intermediate steps
of the calculations.

In spite of the large amount of renewed interest in the BFM,
a regularization independent  proof of renormalizability of gauge
theories quantized with Background Gauge Fixing is still lacking. As a
consequence, there exists no test of compatibility  among the
various regularization schemes allowed with BFM
on different models (see \cite{bkgstd} and \cite{rus}) nor  a verification
of the absence of new anomalies.

This paper is devoted to a regularization independent analysis of the BFM
approach of SU(N) pure Yang-Mills models.
The reader might object that in this situation the analysis is trivial
thanks to Dimensional Regularization compatible with every symmetries of
the model, but we have to meet to this objection that our analysis is
directly extensible to chiral Yang-Mills models
and, up to some little changes, to
the SM. In these latter situations the help of an invariant regularization
fails and one is entitled to fear that the introduction of the
Background fields
could generate new anomalies independent from which are associated with
Quantum Gauge Fields. Therefore our study turns to be necessary. It supplies
a complete answer to the anomaly problem for chiral Yang Mills model and lets
to overcome some difficulties of SM
(the study of a non-semi-simple gauge model in a
spontaneously broken phase is in preparation).

In the first part we study  the model in the tree level approximation.
We want to verify two important features of BFM: the
stability with respect
to the splitting of
the Gauge Field into the  Background Field $V^{a}_{\mu}$ and Quantum Field
$A^{a}_{\mu}$ and
the non-renormalization properties of Background Field  $V^{a}_{\mu}$
(the latter is the usual result \cite{bkg}
that the wave function renormalization  $(V_{0})^{a}_{\mu} =
(Z_{V})^{-\frac{1}{2}} V^{a}_{\mu}$
of this field coincides with the gauge coupling renormalization $g_{0} = Z_{g}
g$:
$$ (Z_{V})^{-\frac{1}{2}} = Z_{g} $$ as
in QED).
We quantize adopting the Background Gauge Fixing
by means of BRS symmetry technique and distinguish at this
level the different role of Quantum Gauge Field and Background
Field: the BRS transformation of the former is the usual covariant derivative
of $\fp$ ghost while the BRS transformation of the latter,
that is naively expected to vanish,  yields a new
BRS invariant ghost for the model.
As it stands, this construction is completely fixed by BRS and
Background Gauge Invariance (and by auxiliary field equations for
ghosts and Lagrange multiplier) and the tree level model is
parametrized by arbitrary constants which can be written
in terms of the  conventional three renormalization constants
fixed by the normalization conditions.

In order to provide a regularization-independent proof of
renormalizability, we work in the BPHZL (\cite{zim},\cite{low})
framework. Avoiding any specific regularization we are allowed to
spoil some symmetries of
the model and, according to the Quantum Action Principle (QAP) (\cite{QAP}),
all sorts of breaking could affect the Slavnov-Taylor (ST) identities, the
Ward Identities (WI) and the auxiliary equations.
As understood (\cite{algeb}), we have to analyse whether
these breakings can be compensated by appropriate finite
non-invariant counterterms. These are
constrained by power counting and by the system of consistency equations
induced by the algebraic properties of the functional operators which
generate the required symmetries on field space.
In particular we search for new possible anomalies and
find that the only possible anomalies are the usual ABJ anomaly
\cite{abj}
which appears both in the ST Identities and in the Ward Identities.

The outline of the paper is as follows. In section 2 we write down the
classical Action and introduce the Background Field
and the Background Gauge Transformations. Then we give a little account of
notation underlining the differences between the conventional
(where the Background Field appears as an infinite collection of gauge
parameters not fixed on a particular classical configuration \cite{bachoo} )
and the gauge-invariant effective action (where the background field is
set equal to the vacuum expectation value of quantum gauge field).
In the same section we define the extended BRS symmetry for this model
and implement it together with all constraints characterising
the tree level action in a system of functional equations.
Section 3 is
devoted to the extension of previous framework to the quantum regime, (\ie the
replacement of the tree level action with the effective action),
and to the analysis of the completeness and
integrability of constraints.
Section 4 is dedicated to the study of the
stability of the tree level model and particular care is
devoted to the stability
of the splitting of gauge field in quantum and classical (background)
parts. In section 5 and in its
subsections we describe the structure of breakings of functional identities
induced
by renormalization and we derive the consistency equations
to which they must satisfied by the breaking terms of the ST.
Then section 6 is completely devoted to
the solution of the consistency equations
by using cohomology techniques  and the Hodge theorem. In
section 7 we check the Ward Identities of Background Gauge Transformations.

\section{The Background Field}

Let us consider the classical field $V^{a}_{\mu}(x)$
belonging to the adjoint representation of Lie algebra of the unitary
group SU(N) and gauge-transforming as:
\beq\label{in1}
V^{a}_{\mu}(x)\rightarrow V^{a}_{\mu}(x)+\nabtil^{ab}_{\mu}\lambda_{b}(x)
\eeq
where $\lambda^{a}(x)$ is the local infinitesimal parameter and
$\nabtil^{ab}_{\mu}$ is the covariant derivative with
respect to $V^{a}_{\mu}$(x).

Unlike the usual point of view we do not take the splitting of the
gauge vector field $A^{a}_{\mu}(x)$
into Background field and Quantum field in {\it a priori} manner, but
we will obtain that as result from the symmetry imposition on the
classical model. Nevertheless we shall call the difference
$Q^{a}_{\mu}=A^{a}_{\mu}-V^{a}_{\mu}$ Quantum gauge field from the beginning.

The Yang-Mills classical action for the gauge
field $A^{a}_{\mu}(x)$  reads:
\beq\label{yap1}
\g^{YM}(A)=\int \dx \left(-\frac{1}{4g^{2}}F^{a}_{\mu\nu}F_{a}^{\mu\nu}\right)
\eeq
where
$F^{a}_{\mu\nu}(A)$ indicates the usual field strength tensor:
\beq\label{yap11}
F^{a}_{\mu\nu}(A)=\de_{\mu}A^{a}_{\nu}-\de_{\nu}A^{a}_{\mu}-
f^{a}\,_{bc}A^{b}_{\mu}A^{c}_{\nu},
\eeq
As it is well known to quantize the model one has to break gauge invariance
by choosing
a gauge-fixing term  and, under the prescriptions of
the corresponding BRS symmetry \cite{bec}, one introduces
the ghost terms.
We classify as Quantum Gauge Transformations (QGT) those which take only the
dynamical degrees of freedom ($A^{a}_{\mu}$) in account
and leave unmodified the
classical background field ($V^{a}_{\mu}$):
\bea\label{yap5} \left\{ \ber{l}
\del A^{a}_{\mu}=\nabla^{ab}_{\mu}\lambda_{b} \\
\del V^{a}_{\mu}=0
\eer \right. \eea

On the contrary, we will call Background Gauge Transformations (BGT)
those where the background field transforms under (\ref{in1}):
\bea\label{yap7} \left\{ \ber{l}
\del A^{a}_{\mu}=\nabla^{ab}_{\mu}\lambda_{b} \\
\del V^{a}_{\mu}=\nabtil^{ab}_{\mu}\lambda_{b}
\eer \right. \eea
so that the difference $Q^{a}_{\mu}$ is a covariant vector:
$\del Q^{a}_{\mu}=f^{a}\,_{bc}\lambda_{b}Q^{a}_{\mu}$

The prescription of Background Field Method (BFM) (\cite{bkg})
tells us how to choose a background-field-dependent
gauge-fixing term $\g^{g.f.}(A,V)$ in order to break the
Quantum Gauge Transformations (\ref{yap5}) but keeping the modified action:
\beq\label{yap33}
\g^{YM}(A) \rightarrow \g_{0}(A,V)=\g^{YM}(A)+\g^{g.f.}(A,V)
\eeq
invariant under Background Gauge Transformations (\ref{yap7}).

The roles of background field and of quantum field
become manifest upon analysing their transformation
properties under (\ref{yap5}) and (\ref{yap7}).
With respect the former the background field stays unchanged and. Therefore
it does not contribute to the conserved current associated to the
rigid symmetry
and coupled to the  vector field $A^{a}_{\mu}$. The latter
transformations will be implemented in the quantum theory by BRS symmetry.
Under  (\ref{yap7}) the field $Q^{a}_{\mu}(x)$
behaves as a vector of the adjoint representation and it contributes to the
conserved current of background symmetry for which the
background field $V^{a}_{\mu}$ acts as a classical external source.

We express this last symmetry of the action $\g_{0}(A,V)$
in the functional language by the Ward Identities (WI):
\beq\label{in71}
W^{a}(x)\g_{0}(A,V) \equiv
-\left(\nabla^{ab}_{\mu}\dfa{A}{b}{\mu}{(x)}+\nabtil^{ab}_{\mu}
\dfa{V}{b}{\mu}{(x)}\right)\g_{0}(A,V)=0.
\eeq
The local functional:
\beq\label{in72}
\tilde{\g}_{0}(A)=\g_{0}(A,V)|_{A=V}
\eeq
then satisfies the following Ward Identities:
\beq\label{in731}
\left[W^{a}(x)\g_{0}(A,V)\right]_{A=V}=
-\nabla^{ab}_{\mu}\dms{\dfa{A}{b}{\mu}{(x)}} \tilde{\g}_{0}(A)=0.
\eeq

Now we are ready to make practical our preliminary considerations.
We extend the field space of the theory introducing the ghost fields
$\om^{a}(x)$, $\no^{a}(x)$ which have naive dimension one and carry
Faddeev-Popov $\q$ charge 1 and -1 respectively. We also introduce
the Nakanishi-Launtrup Lagrange
multiplier $b^{a}(x)$ (\cite{bec}, \cite{ko})
with dimension two and with no $\q$ charge.
Each of these fields transforms as a vector of the adjoint
representation under BGT.

On this set of fields we can define the conventional
BRS transformations by
\bea\label{in12} \left\{ \ber{l}
\vspace{.3cm}
s A^{a}_{\mu}=\nabla^{ab}_{\mu} \om_{b} \\
\vspace{.3cm}
s \om ^{a}=\frac{1}{2}f^{a}~\!_{bc} \om^{b} \om^{c} \\
\vspace{.3cm}
s \no ^{a}=b^{a} ~~~~~ s b^{a}=0
\eer \right. \eea
where we choose to indicate with $s$
the generator of BRS.

Then, as usual \cite{bec}, we complete the set of fields by introducing
the external sources
$\gamma^{a}_{\mu}(x),\zeta^{a}(x)$
(respectively with $\q$ charge -1 and -2 and both
with dimension two) coupled respectively to BRS
variations of the gauge field $A^{a}_{\mu}(x)$ and ghost $\om^{a}(x)$.
Again, these latter fields are covariant under BGT, invariant under
BRS transformations and allow us to translate the non linear BRS
symmetry of the model into
the Slavnov-Taylor (ST) identities:
\beq\label{slta}
s(\g_{0})=\dms{\int} \dx \left[
\fd{\g_{0}}{\gamma}{a}{\mu} \fd{\g_{0}}{A}{\mu}{a} +
\fdd{\g_{0}}{\zeta}{a} \fdu{\g_{0}}{\om}{a} +
b^{a} \fdu{\g_{0}}{\no}{a}
\right]=0
\eeq

In the same way we can introduce
the Ward-Takahashi Identities by means of the generator of BGT $W(\lambda)$:
\bea\label{in13} \ber{c}
W(\lambda)\g_{0}[A,V,\om,\no,b,\gamma,\zeta] \equiv
\dms{\int} \dx \left[
(\nabla\lambda)^{b}_{\mu}\fd{\g_{0}}{A}{b}{\mu} +
(\nabtil\lambda)^{b}_{\mu}\fd{\g_{0}}{V}{b}{\mu} +
f^{a}~\!_{bc}\lambda^{b}\Phi^{c}\fdu{\g_{0}}{\Phi}{a}
\right]=0
\eer \eea
for which $\Phi^{a}(x)$ collects all covariant fields
$\om^{a}(x),\ \no^{a}(x),\ b^{a}(x)$
$\gamma^{a}_{\mu}(x)$ and $ \zeta^{a}(x)$.
It is easy to note that the two set of transformations, BRS and BGT
commute, so we have the possiblity
to build a model invariant under both.

Now we can write down our choice of gauge-fixing by using directly the
functional equation for the Lagrange multiplier $b^{a}$ :
\beq\label{id28}
\dms{\int} \dx \bet_{a} \left[ \fdu{\g_{0}}{b}{a} +
\nabtil^{ab}_{\mu}(A_{b}^{\mu}-V_{b}^{\mu})-\alp b^{a} \right]=0
\eeq
where $\bet_{a}(x)$ is an arbitrary test function covariant under BGT and
$\alp$ is the gauge parameter.

Because of the nilpotence of BRS transformations we build the
quantizable version of the theory by adding the BRS variation of an
integrated polynomial in the fields
to the invariant action  $\Gamma^{YM}[A]$ (\ref{yap1}):
\beq\label{gf}
\g_{0}[A,V,\om,\no,b,\gamma,\zeta]=\Gamma^{YM}[A]+ s
\int \dx
\left[
\no_{a}\left( \nabtil ^{ab}_{\mu}(A-V)^{\mu}_{b}+\frac{\alp}{2}b^{a} \right)+
\gamma^{a}_{\mu}A_{a}^{\mu}+\zeta^{a}\om_{a}
\right]
\eeq

In this way the dependence of background field is confined
in the gauge-fixing and ghost
terms (this feature suggests that physical observables of
this theory are independent on the background field \cite{ags}). To
implement this feature in the functional language we take the derivative
of (\ref{gf}) with
respect to the background field $V^{a}_{\mu}(x)$, obtaining:
\beq\label{id4}
\dms{\fd{\g_{0}}{V}{a}{\mu}} =
s \left(\nabla^{ab}_{\mu}\no_{b}\right)
\eeq
In the second member of (\ref{id4}) the  BRS variation
of a composite operator appears.  Then,
to freely translate this equation in functional form, we have to couple to it
a new anticommuting field  $\Omega^{a}_{\mu}(x)$ with
dimension two, positive $\q$ charge and
invariant under BRS:
\beq\label{ido6}
\Gamma_{0} \rightarrow \Gamma_{0}+
\int \dx \left(\Om^{a}_{\mu} \nabla^{ab}_{\mu}\no_{b}\right)
\eeq
thus  we can lead back
the background field equation (\ref{id4}) to a generalised
ST identities:
\beq\label{ido7}
s(\g_{0}) =
-\dms{ \int \dx \left( \Om^{a}_{\mu} s\nabla^{ab}_{\mu}\no_{b} \right)}=
-\dms{ \int \dx \left( \Om^{a}_{\mu} \fd{\g_{0}}{V}{a}{\mu} \right)}
\eeq
by differentiating with respect to the field $\Om^{a}_{\mu}(x)$ and set
it to zero.
As we will see in the following, the background field equation (\ref{id4})
is fundamental to control  the splitting for the gauge field
into background gauge field and the quantum one.

It is immediate to note that the previous equation (\ref{ido7}) can be
simply obtained extending the BRS transformations (\ref{in12}) to the
background field:
\bea\label{in123} \left\{ \ber{l}
\vspace{.3cm}
s V^{a}_{\mu}= \Om^{a}_{\mu} \\
s \Om^{a}_{\mu} = 0
\eer \right. \eea
here $\Om^{a}_{\mu}$ assumes the role of a ghost for the background field,
and then, differentiating the (\ref{ido7}) with respect to
the ghost $\Om^{a}_{\mu}$ we get:
\beq\label{ciao}
\fd{\g_{0}}{V}{a}{\mu}  = s \left[ \fd{\g_{0}}{\Om}{a}{\mu} \right]
\eeq
\ie the dependence of the theory on the background field is completely confined
to  a BRS variation.

\section{Quantum theory and Functional Identities}
We now pass to the quantum theory, defining the generating functional
$Z[\bar{J},\bar{\eta},V]$ by means of the Feynman integral:
\beq\label{id7}
Z[\bar{J},\bar{\eta},V]=
\int {\cal D}(A){\cal D}(\om){\cal D}(\no){\cal D}(b)
e^{(i\Gamma_{0}[\Psi,V,\bar{\eta}]+S.T.)}
\eeq
where we indicate with S.T. the source terms
and $\bar{J}=\{J^{a}_{\mu},\bar{\xi}^{a},\xi^{a},Y^{a}\}$ collects
all the sources for the quantum fields
$\Psi=\{A^{a},\om^{a},\no^{a},b^{a}\}$ and
$\bar{\eta}=\{\gamma^{a}_{\mu}(x),\zeta^{a}(x),\Om^{a}_{\mu}(x)\}$ for
the external fields.  We explicitly render the dependence of
the generating functional $Z[\bar{J},\bar{\eta},V]$ on the
background field distinguishing the sources $\bar{J}$
form the external fields $\bar{\eta}$.
We define the effective action by the Legendre transformation \cite{MII}:
\beq\label{id27}
\Gamma[\Psi,\bar{\eta},V]_{\Psi=\frac{\delta Z_{c}}{\delta \bar{J}}}
=Z_{c}[\bar{J},\bar{\eta},V,]-\int \dx
(J^{a}_{\mu}A_{a}^{\mu}+\no^{a}\xi_{a}+\bar{\xi}^{a}\om_{a}+b^{a}Y_{a})
\eeq

We want to point out that the derivation of the following equations
for generating functional is pure formal because we did not keep in
account the unavoidable regularization procedure. For the moment
we just suppose
to use regularised quantities preserving all sorts of symmetry of
our model, although, as known in the Standard Model,
no regularization procedure actually does it. We postpone
the study of the effects of regularization of Feynman integrals
in the next section and here we simply translate
the Lagrange multiplier equation, the WT Identities
and the ST Identities in terms of the generating   functional (\ref{id7}).

We find for the Lagrange multiplier equation (\ref{id28}) the
following relation:
\bea\label{id9}  \ber{l}
E(\bet)Z[\bar{J},\bar{\eta},V] \equiv \\~~
\\
\equiv \dms{\int} \dx \bet_{a}(x)
\left[ \nabtil^{ab}_{\mu}\left(-i\dms{\dfa{J}{b}{\mu}{(x)}}-
V^{b}_{\mu}(x) \right)
-i\alp\dms{\dfu{Y}{a}{(x)}}+Y^{a}(x) \right] Z[\bar{J},\bar{\eta},V]=0.
\eer \eea
In the same way we translate the ST identities (\ref{ido7})
into  the linear equation:
\beq\label{id15}
{\cal S}Z[\bar{J},\bar{\eta},V]\equiv \int \dx
\left[J_{a}^{\mu}\dfa{\gamma}{a}{\mu}{(x)}+
\Om_{a}^{\mu}\dfa{V}{a}{\mu}{(x)}-\bar{\xi}_{a}\dfu{\zeta}{a}{(x)}
-\xi_{a}\dfu{Y}{a}{(x)}
\right]Z[\bar{J},\bar{\eta},V]=0.
\eeq
Due to  nilpotence of BRS transformations
and commutation properties of the external field $\Om^{a}_{\mu}(x)$
and of background field $V^{a}_{\mu}(x)$,
the functional operator ${\cal S}$ is nilpotent.

As usual \cite{bec} we recover the Faddeev-Popov field equation calculating
explicitly the commutator between the Slavnov-Taylor operator ${\cal S}$
and the operator $E(\bet)$ defined in (\ref{id9}):
\bea\label{id21} \ber{l}
\Sigma(\bet) Z[\bar{J},\bar{\eta},V]\equiv
\left[{\cal S},E(\bet) \right]Z[\bar{J},\bar{\eta},V]= \\~~\\
=\dms{\int} \dx
  \bet_{a}(x)\left[
	\left(
      \de^{\mu}\Omega^{a}_{\mu} -if^{a}~\!_{bc}\Omega^{b}_{\mu}
	 \dms{\dfa{J}{c}{\mu}{(x)}}
	\right)
	+\xi^{a}+i\nabtil^{ab}_{\mu}\dms{\dfa{\gamma}{b}{\mu}{(x)}}
  \right]   Z[\bar{J},\bar{\eta},V]=0
\eer \eea

Finally taking into account the invariance properties of
the classical action $\g_{0}$ and the variation of source terms in
the generating functional (\ref{id7}), we find
the corresponding WT Identities for
$Z[\bar{J},\bar{\eta},V]$:
\bea\label{id19} \ber{l}
W(\lambda)Z[\bar{J},\bar{\eta},V]\equiv \\~~\\
\!\! \dms{\int} \dx \!\left[
J^{\mu}_{a} \left( \de_{\mu} \lambda^{a}+if^{a}_{bc} \lambda^{b}
\dms{\dfa{J}{c}{\mu}{(x)}} \right)+
\nabtil ^{ab}_{\mu} \lambda_{b}\dms{\dfa{V}{a}{\mu}{(x)}}+
i(f^{a}_{bc}\lambda_{a})\Phi^{b}\dms{\dfu{\Phi}{c}{(x)}}
\right]Z[\bar{J},\bar{\eta},V]=0
\eer \eea
where we collected all covariant fields $\om^{a},\no^{a},\dots$ under
a single symbol $\Phi^{a}(x)$.

The complete algebra of functional operators
$\{{\cal S},W(\bet),\Sigma(\bet),E(\bet) \}$ is given by the commutators:
\beq\label{comm1}
[W(\lambda),E(\bet)]=E(\lambda\wedge\bet)
{}~~~~~[W(\lambda),W(\bet)]=W(\lambda\wedge\bet)
{}~~~~~[E(\lambda),\Sigma(\bet)]=0
\eeq
\beq\label{comm2}
[{\cal S},E( \bet )]=\Sigma ( \bet ) ~~~
\{ { \cal S},\Sigma( \bet ) \} = 0 ~~~[{ \cal S},W(\bet)]=0
{}~~~[W(\lambda),\Sigma(\bet)]=\Sigma(\lambda\wedge\bet)
\eeq
i.e. they belong to an involutive algebra
and, by Fr\"obenius theorem, they generate a
completely integrable differential system.

\section{Stability of classical model}
We now come back to the functional equations for the classical action
$\g_{0}$. In the previous sections
we have defined a classical action and we have described the
set of constraints to which this must satisfy.We now want to verify that
the most general solution of these constraints is our starting classical
action up to possible multiplicative field renormalizations.  Then
in the following we shall
consider the local functional $\g_{0}$ unknown and we proceed
to solve completely the set of functional equations.

The first step toward a simplification of the cumbersome set of
functional equations satisfied by $\g_{0}$  consists in substituting it with
the new modified version:
\beq\label{id32}
\gh_{0}=\g_{0} - \int \dx
\left[b^{a}\nabtil^{ab}_{\mu}(A_{b}^{\mu}-V_{a}^{\mu})+
\dms{\frac{\alp}{2}}b^{a}b_{a}+\Omega_{a}^{\mu}\nabla^{ab}_{\mu}\no_{b}\right]
\eeq
We subtracted the gauge-fixing terms because of their
non-renormalization properties and we also chose to subtract away
the $\Om$-dependent terms.

Then it is easy to verify that the new effective action $\gh_{0}$
satisfies the system of equations:
\bea\label{newsys} \left\{  \ber{l}
\oop{.3cm}
\dms{\int}\dx \bet^{a}\dms{\fdu{\gh_{0}}{b}{a}}=0 \\
\oop{.3cm}
\dms{\int} \dx \bet_{a} \left(\fdu{\gh_{0}}{\no}{a} -
\nabtil^{ab}_{\mu}\fd{\gh_{0}}{\gamma}{b}{\mu} \right) = 0 \\
\oop{.3cm}
\dms{\int} \dx \left[
\dms{\fd{\gh_{0}}{A}{a}{\mu}}  \dms{\fd{\gh_{0}}{\gamma}{\mu}{a}} +
\dms{\fdu{\gh_{0}}{\om}{a}} \dms{\fdd{\gh_{0}}{\zeta}{a}} +
\Om^{a}_{\mu} \left( \dms{\fd{\gh_{0}}{V}{a}{\mu}- f_{abc}
\no^{b} \fd{\gh_{0}}{\gam}{c}{\mu}}\right) \right]=0          \\
\dms{\int}\dx \left[
\nabla_{\mu}^{ab} \lambda^{b} \dms{\fd{\gh_{0}}{A}{c}{\mu}} +
\nabtil ^{ab}_{\mu} \lambda_{b}\dms{\fd{\gh_{0}}{V}{a}{\mu}} +
(f^{a}~\!_{bc}\lambda_{a})\Phi^{b} \dms{\fdu{\gh_{0}}{\Phi}{c}}
\right](x) = 0
\eer \right. \eea
The first equation shows that the new action $\gh_{0}$ does not dependent upon
the  $b^{a}$ field; the second one gives an important constraints on the
antighost  field $\no^{a}$, forcing it to appear only in the linear
combination:
\beq\label{id361}
\hg ^{a}_{\mu}=\gamma^{a}_{\mu}-\nabtil^{ab}_{\mu}\no_{b}
\eeq
Since every change from $\g_{0}$ to $\gh_{0}$
is gauge invariant no breaking term appears in the
WT Identities (\ref{id19}). On the other hand we have to modify
slightly the ST identities.

Because $\hg^{a}_{\mu}(x)$ is of dimension two and has  charge $\q$=-1 the
action $\gh_{0}$  may only depend linearly upon it.
Then we can rewrite the ST identities  (\ref{newsys})
in the form:
\beq\label{id40}
{\cal D}_{\gh_{0}}\gh_{0} \equiv
\dms{\int} \dx \left[
\dms{\fd{\gh_{0}}{A}{a}{\mu}}  \dms{\fd{\gh_{0}}{\hg}{\mu}{a}} +
\dms{\fdu{\gh_{0}}{\om}{a}} \dms{\fdd{\gh_{0}}{\zeta}{a}} +
\Om^{a}_{\mu} \dms{\fd{\gh_{0}}{V}{a}{\mu}} \right]=0
\eeq

In  order
to simplify the analisys of ST identities we introduce
the linear operator:
\bea\label{id42} \ber{l}
{\cal F}_{\gh_{0}}\equiv
\dms{\int} \dx \left[
\dms{\fd{\gh_{0}} {A}{a}{\mu}} \dms{\fdf{\hg}{\mu}{a}} +
\dms{\fd{\gh_{0}}{\hg}{\mu}{a}} \dms{\fdf{A}{a}{\mu}} +
\dms{\fdu{\gh_{0}}{\om}{a}} \dms{\fdfu{\zeta}{a}} +
\dms{\fdd{\gh_{0}}{\zeta}{a}} \dms{\fdfu{\om}{a}} +
\Omega^{a}_{\mu}\dms{\fdf{V}{a}{\mu}} \right]
\eer \eea
The ST identities themselves imply the consistency equations
\beq\label{id44}
{\cal F}_{\gh_{0}}{\cal D}_{\gh_{0}}\gh_{0}=0
\eeq
and the commutability of BGT with BRS
transformations may be rewrited in the form:
\beq\label{id451}
{\cal F}_{\gh_{0}}W(\lambda)-W(\lambda){\cal D}_{\gh_{0}}=0
\eeq

We now can complete the check of  stability on the classical model by
finding the general solution of functional equations (\ref{newsys})
for the functional $\gh_{0}$ which depends only on the reduced set of fields
$A^{a}_{\mu},\hg^{a}_{\mu},\om^{a},\zeta^{a}$
and $\Om^{a}_{\mu}$.

We start analysing the $\Om^{a}_{\mu}$ dependent part of the action.
Since this field is characterized by  $\q$ charge  +1
and naive dimension 2, the only gauge invariant integrated
polynomial which respects
the quantum numbers of the functional $\gh_{0}$ is forced to
be
\beq\label{so2}
\Delta=u\int \dx (\Om^{a}_{\mu}\hg_{a}^{\mu})
\eeq
where u is a constant.

Redefining a new action with
\beq\label{so1}
\gl_{0}= \gh_{0} - u\int \dx \left( \Om^{a}_{\mu}\hg_{a}^{\mu} \right)
\eeq
and exploiting the independence of new action $\gl_{0}$ on the $\Om^{a}_{\mu}$
field,
the ST identities (\ref{id40}) split into the system of functional
equations:
\bea\label{so4} \left\{ \ber{l}
\oop{.3cm}
\dms{\int} \dx \left[
\dms{\fd{\gl_{0}}{A}{a}{\mu}} \dms{\fd{\gl_{0}}{\hg}{a}{\mu}} +
\dms{\fdu{\gl_{0}}{\om}{a}} \dms{\fdd{\gl_{0}}{\zeta}{a}}
\right]=0 \\
\oop{.3cm}
\left(
\dms{\fd{\gl_{0}}{V}{a}{\mu}} - u \dms{\fd{\gl_{0}}{A}{a}{\mu}}
\right) = 0.
\eer \right. \eea

{}From the second equation we deduce that the vector fields $A^{a}_{\mu}(x)$
and
$V^{a}_{\mu}(x)$ appear in the solution of our problem only through the
linear combination:
\beq\label{so6}
\h{A}^{a}_{\mu}=A^{a}_{\mu}+uV^{a}_{\mu}
\eeq
If we substitute the field $\h{A}^{a}_{\mu}(x)$ in place of
$A^{a}_{\mu}(x)$, the first equation of the system (\ref{so4}) coincides
with the usual ST identities, for which the solution is
well-known \cite{bec}:
\beq\label{so8}
\gl _{0}[\h{A},\om,\zeta,\hg]=
\dms{\int} \dx \left[-\frac{1}{4g^{2}}F^{\mu\nu}_{a}F^{a}_{\mu\nu}+
\frac{z}{2}f_{abc}\zeta ^{a}\om ^{b}\om ^{c}+
(zZ)\hg ^{a}_{\mu}(\de _{\mu}\om^{a}-
Z^{-1}f_{abc}\h{A}^{b}_{\mu}\om^{c})
\right]
\eeq
where
\beq\label{so9}
F^{a}_{\mu\nu}\h{A})=
Z^{-1}(\de_{\mu}\h{A}^{a}_{\nu}-\de_{\nu}
\h{A}^{a}_{\mu}
-Z^{-1}f^{a}\,_{bc}\h{A}^{b}_{\mu}\h{A}^{c}_{\nu})
\eeq
and the constants $g,~ z$ and $Z$ are the only free parameters.

As it stands the linear combination (\ref{so6}) is not background gauge
invariant, but requiring this property we arrive to:
\beq\label{so11}
W(\lambda)(Z^{-1}\h{A}^{a}_{\mu})=Z^{-1}(1+u)\de_{\mu}\lambda^{a}-
f^{a}\,_{bc}\lambda^{b}Z^{-1}\h{A}^{c}_{\mu}
\eeq
\ie  $Z=1+u$.
Rewriting this result in terms of $Q^{a}_{\mu}$ and $V^{a}_{\mu}$ we
obtain
\beq\label{so13}
Z^{-1}\h{A}^{a}_{\mu}=Z^{-1}(A+uV)^{a}_{\mu}=Z^{-1}Q^{a}_{\mu}+V
\eeq
from which one immediately sees that only the $Q^{a}_{\mu}$ field is
multiplicative renormalized.
Then for a Yang-Mills theory quantized with background gauge the splitting
between quantum field $Q^{a}_{\mu}$ ( it is now  manifest why we called it
Quantum Field) and the background field $V^{a}_{\mu}$ is obtained from
the constrains of symmetry imposed on it. This also shows that the
decomposition is stable and no radiative correction may
mix the quantum part with classical one. We wish to stress that
the field renormalization $\h{A}\rightarrow Z^{-1}\h{A}$ is
the usual field renormalization of Yang-Mills theory.

We conclude this part with some considerations about the
normalization conditions. The manifest gauge invariance of the effective
action (\ref{in72}) implies the counter-terms dependence only upon
gauge-independent ``physical renormalization'' \ie on charge renormalization.
Then we have no need to specify the normalization conditions for quantum fields
\ie for the wave function renormalization of quantum gauge fields
and of the ghost fields, or in other words, the physical quantities are
insensitive on these intermediate normalization conditions.
This aspect of BFM will be very useful in the SM in the spontaneously
symmetry broken phase because of large amount of gauge-dependent (``non
physical'') renormalization constants especially if one decides to work
in a covariant gauge.

In the parametrization of the tree level action given in (\cite{gaude})
the arbitrariness is confined by constraints
to the wave function renormalizations
$Z$ for gauge fields and $z$ for ghost fields and in an overall renormalization
of the action.
In fact in this parametrization it is easy ( by means of an
enlarged set of BRS transformations comprising the BRS variation for
the gauge parameter) to show the
gauge-independence of the overall renormalization. On the contrary the
two constant $Z,z$ result to be gauge-dependent parameters, and so are
the corresponding anomalous dimensions $\gam_{A},\gam_{c}$.

This parametrization is particular advantageous in the BFM because
here the overall constant coincides with the inverse of the square of the
coupling constant, so that the overall renormalization for
the effective action $\g[A,V]$ becomes the overall renormalization
for the gauge invariant effective action $\widetilde{\g}[A]$.
Gauge invariance then tell us that this renormalization is the charge
renormalization.

Then we can fix the following normalization condition:
\bea\label{norcond}  \ber{c}
\oop{.4cm}
\widetilde{\g}_{A^{a}_{\mu}A^{b}_{\nu}}(p) = \del^{ab}
\left(g_{\mu\nu}-\dms{\frac{p_{\mu}p_{\nu}}{p^{2}}}\right) \Lambda_{A}(p) \\
\dms{\frac{\de}{\de p^{2}}} \Lambda_{A}(p) \left. \right|_{p^{2} =
\kappa^{2}} =\frac{-1}{g^{2}}
\eer \eea
where $\kappa^{2}$ is the (euclidean) normalization point.
This choice ok $\kappa^{2}$ is irrelevant at the
tree level, and in any case under complete control by means of
Callan-Symanzik equation.

The particular parametrization chosen here for the BFM-version of pure
Yang-Mills
implies immediately the Kallosh theorem \cite{kallo}. In fact the
independence of the overall renormalization from gauge parameter is
equivalent to that of the coupling constant renormalization, \ie
the Kallosh theorem.  This fact implies also the independence from
a fixed gauge parameter for the normalization condition (\ref{norcond})
(compare this with the corresponding normalization condition in \cite{gaude}).

\section{Renormalization}
\subsection{Consistency equations}
Since the symmetries characterising the theory are acceptable for any purpose,
including for instance the background gauge invariance of the quantum action
$\gh$ and the determination of counter terms, it is necessary that they
survive the quantization process.
In this section our intent is to describe how the system of previous equations
for the functional $\gh$ will be modified by the breaking terms induced by
radiative corrections and by the subtraction procedure and we will
verify that the system can be restored order by order by an appropriate
choose of counter terms.

The main tool for this investigation is
the Quantum Action Principle (QAP) \cite{QAP},  proved
in the BPHZL renormalization scheme
\cite{zim}, \cite{low}, which states that the functional equations of
the model get broken at the quantum level by local integrated insertions.
Moreover, as it is usual \cite{bec} in the standard algebraic procedure,
the commutation properties (\ref{comm1}) and (\ref{comm2})
of functional operators and the features of local insertions allow us
to derive a
set of consistency equations to which these breaking terms have to satisfy.

Solving the set of consistency equations we individuate all possible
breaking terms. In general, at this point
two situations could happen: either all
these breaking terms are variations of non symmetric counter terms for the
tree level action,\ie they can be compensated,
or there is at least a non-compensable breaking terms; in
this latter case the corresponding broken equation gets an anomalous term or,
on the other hands, the symmetry described by that equation is not a
symmetry any longer at the quantum level.

In the present section we use the notation
$\gh$ to indicate 1PI generating functional, the
symbol $\gh_{0}$ for the corresponding tree approximated action,
and $\gh_{Eff}$ for the effective action where the coefficients of
integrated field polynomial of
$\gh_{0}$ are replaced by formal power series of $\hbar$. The
last functional $\gh_{Eff}$ deals with order by order counter terms
that implement the normalization conditions and symmetries at the
quantum level.

In order to protect our calculations from infrared divergences
induced by a zero momentum subtractions of BPHZ scheme, we adopt the
modification introduced by J.Lowenstein \cite{low}:
\bea\label{con7} \ber{l}
\gh_{0}(Q,V,\om,\hg,\zeta,\Omega)\rightarrow \\
\hspace{.5cm} \rightarrow \gh_{0}(Q,V,\om,\hg,\zeta,\Omega;s)\equiv
\gh_{0}(Q,V,\om,\hg,\zeta,\Omega)+
\mu^{2}(s-1)^{2}\dms{\int} \dx
\left(\frac{1}{2}Q^{a}_{\mu}Q_{a}^{\mu}+\no^{a}\om_{a} \right)
\eer \eea
The propagators for the massless fields $Q^{a}_{\mu}$, $\om^{a}$ e $\no^{a}$
acquire a parameter s dependent mass; this parameter s varies between zero
and one and will play the role of an additional subtraction variable (like
an external momentum p). At the end of calculations it is to be put equal
to one for all massless fields. At $s\neq1$ such mass term will describe a
massive field (with off-shell normalization conditions \cite{low}).

Nevertheless we have to note that the ST identities  (\ref{slta})
for the classical action $\gh_{0}$ is spoiled by those new terms because
they are not BRS invariant, and, in the same way, also the
equation of motion for the ghost field $\no^{a}(x)$ is modified.
The presence of these breaking terms does not alter the content
of BRS symmetry because they are  only soft breaking terms
and they have to be treated by the technique of the reference \cite{clrk},
\cite{abj} and \cite{algeb}. By the way we will find some other
(s-1)-dependent terms during the building of counter terms, but we have to
keep in mind the limit $s\rightarrow 1$ as final act of renormalization
procedure \cite{clrk} and so we have not to show their explicit structure.
On the other way, we note
that the mass terms are background gauge invariant, and no breaking term
occurs for the corresponding WT Identities.

Also for ghost equation of motion we meet the same difficulties of
ST identities because of regularization of massless
propagators, and the basic simple constrain given by this equation is lost.
Nevertheless, since we are interested into the theory for $s=1$, we assume
the dependence of the functional $\gh$ on fields
$\gam^{a}_{\mu}(x)$ and $\no_{b}(x)$  is only through the linear
combination:
\beq\label{con11}
\hg^{a}_{\mu}=\gam^{a}_{\mu}-\nabtil^{ab}_{\mu}\no_{b}
\eeq
which is correct up to $(s-1)$ corrections. (See \cite{henri} for a more
detailed discussion).

We now can complete the translation of our functional identities in the
BPHZL contest obtaining:
\bea\label{con4} \left\{ \ber{l}
\vspace{.3cm}
\dms{\int} \dx \left[\dms{ \fd{\gh}{A}{a}{\mu}}
\dms{\fd{\gh}{\hg}{\mu}{a}} +
\dms{\fdu{\gh}{\om}{a}} \dms{\fdd{\gh} {\zeta}{a}}
+\Om^{a}_{\mu}\dms{\fd{\gh}{V}{a}{\mu}} \right]=
\Delta_{ST}\gh
 \\
\dms{\int} \dx \left[\nabtil^{ab}_{\mu}\lambda_{b}
\dms{\fd{\gh}{V}{a}{\mu}}  +
\nabla^{ab}_{\mu}\lambda_{b}
\dms{\fd{\gh}{A}{a}{\mu}} +
f^{a}~\!_{bc}\lambda_{a}\Phi_{b}\dms{\fdu{\gh}{\Phi}{c}} \right]=
\Delta_{W}\gh
\eer \right. \eea

These equations are different from the classical ones
(\ref{slta}) and (\ref{in13}) because of the breaking terms
$\Delta_{ST}\gh$ and $\Delta_{W}(\lambda)\gh$ which sum up the non
invariant contributions from loop computations.
Following the notations of \cite{clrk} or \cite{zim} we can
make the UV, IR and $(s-1)$ indices of subtraction for
the breakings $\Delta_{ST}$ e $\Delta_{W}(\lambda)$ explicit:
\bea\label{con13} \ber{l}
\vspace{.1cm}
\Delta_{ST}\gh=\dms{\int} \dx N^{5}_{5,2}[Q_{ST}(x)]\gh \\
\Delta_{W}(\lambda)\gh=
\dms{\int} \dx \lambda_{a}N^{4}_{4,2}[Q^{a}_{W}(x)]\gh
\eer \eea
where the first local polynomial $Q_{ST}(x)$ is scalar under
Poincar\'e transformations, invariant under rigid gauge transformations and
carries the charge $\q$=1; the second one: $Q^{a}_{W}(x)$ is
also scalar under Poincar\'e transformations, carries an index of
adjoint representation of rigid gauge transformations and no $\q$ charge.
These quantum numbers assigned to them classify completely the algebraic
structure of the breaking terms. Their
coefficients have to be fixed on normalization conditions.

We now proceed by induction technique supposing that breakings of the
identities (\ref{con4}) were reabsorbed up to the $\hbar^{n-1}$-order of
perturbation theory and, using the properties of
Zimmermann\'s normal products, we can separate
$\hbar^{n}$-order contributions from those of higher order:
\bea\label{con5} \ber{l}
\vspace{.5cm}
\Delta_{ST}\gh=\hbar^{n}\dms{\int} \dx P^{(n)}_{ST}(x)+O(\hbar^{n+1}) \\
\Delta_{W}(\lambda)\gh=\hbar^{n}\dms{\int} \dx \lambda_{a} P^{(n)a}_{W}(x)+
O(\hbar^{n+1})
\eer \eea

{}From these definitions  we deduce the consistency equations to
which the $\Delta_{ST}\gh$ and $\Delta_{W}(\lambda)\gh$ have to
satisfy. We rewrite the system (\ref{con4}) in the synthetic notation:
\bea\label{con18} \left\{ \ber{l}
\vspace{.5cm}
\D\gh=\hbar^{n}\Delta^{(n)}_{ST}+O(\hbar^{n+1}) \\
W(\lambda)\gh=\hbar^{n}\Delta^{(n)}_{W}(\lambda)+O(\hbar^{n+1})
\eer \right. \eea
and recall the nilpotence or the commutation properties of
functional operators $\D,\F$ and $W(\lambda)$ (\ref{comm1}) and (\ref{comm2}):
\beq\label{con19}
\F\D\gh=0, ~~~\F W(\lambda)-W(\lambda)\D=0, ~~~
[W(\lambda),W(\bet)]=W(\lambda\wedge\bet)
\eeq
where $\F$ is the linear ST operator of the previous section.

If we act with the operator $\F$ on the left upon the first
equations of (\ref{con18}) and by (\ref{con19}), we obtain:
\beq\label{st9}
\F\D\gh=\hbar^{n}\F\Delta^{(n)}_{ST}+O(\hbar^{n+1})=0
\eeq
from which, organising at the best the $\hbar^{n}$-order terms,
and recalling $\gh=\gh_{0}+\hbar\gh_{1}$, it follows:
\beq\label{st10}
\hbar^{n}\F\Delta^{(n)}_{ST}+O(\hbar^{n+1})=
\hbar^{n}\oF \Delta^{(n)}_{ST}+O(\hbar^{n+1})
\eeq
\ie the consistency equations:
\beq\label{con24}
\oF \Delta^{(n)}_{ST}=0
\eeq

On the other hand acting respectively on the first and on the second
equation with the operators $W(\lambda)$ and $\F$, we arrive at:
\bea\label{con26} \left\{ \ber{l}
\vspace{.5cm}
W(\lambda)\D\gh=\hbar^{n}W(\lambda)\Delta^{(n)}_{ST}+O(\hbar^{n+1}) \\
\F W(\lambda)\gh=\hbar^{n}\oF \Delta^{(n)}_{W}(\lambda)+O(\hbar^{n+1})
\eer \right. \eea
and by the (\ref{con19}) at:
\beq\label{con27}
W(\lambda)\Delta^{(n)}_{ST}-\oF \Delta^{(n)}_{W}(\lambda)=0
\eeq
up to the order $\hbar^{n}$.

Finally  the commutation properties for the
$W(\lambda)$ operators imply the following consistency equations:
\beq\label{con30}
W(\lambda)\Delta^{(n)}_{W}(\bet)-W(\bet)\Delta^{(n)}_{W}(\lambda)=
\Delta^{(n)}_{W}(\lambda\wedge\bet)
\eeq
which can be identified
with the well-known Wess-Zumino conditions \cite{wee}.

In the next sections we will completely analyse the set of consistency
equations (\ref{con24}), (\ref{con27}) and (\ref{con30}).
We want to recall that also the Faddeev-Popov ghost equations and
the Lagrange multiplier equations are modified by
breaking terms but for their analysis we remand the reader to
\cite{henri} again.

\subsection{Consistency equations for ST breakings}

To analyse the consistency equations  (\ref{con24})
for ST breakings $\Delta^{(n)}_{ST}$
we have to decompose the operators $\F$ and $\D$ in simpler ones:
\bea\label{st5} \ber{l}
\vspace{.5cm}
\B= \dms{\int} \dx \left(
\dms{\fd{\gh}{A}{a}{\mu}  \fdf{\hg}{\mu}{a} +
\fdu{\gh}{\zeta}{a} \fdfd{\om}{a}} \right) ~~~~
\aB_{\gh} = \dms{\int} \dx
\left(\dms{\fd{\gh}{\hg}{\mu}{a} \fdf{A}{a}{\mu} +
\fdd{\gh}{\om}{a}  \fdfu{\zeta}{a}} \right) \\
\aR= \dms{\int} \dx \left(\Omega^{a}_{\mu} \dms{\fdf{V}{a}{\mu}}
\right)
\eer \eea
by means of them we have:
\beq\label{st4}
\vspace{.2cm}
{\cal D}_{\gh}=\DD+\aR=\meta (\B+\aB_{\gh})+\aR ~~~~
{\cal F}_{\gh}=2\DD+\aR=(\B+\aB_{\gh}+\aR)
\eeq
where the $\gh$-dependent and $\gh$-independent
parts of operators $\D$ and $\F$
are splitted.

We now want to show how to build the counter terms.
One easily checks that the following algebraic relations hold:
\beq\label{st6}
\B\gh=\paB_{\gh}\gh~~~\B\Delta=\paB_{\Delta}\gh~~~
\paB_{\gh}\Delta=\pB_{\Delta}\gh
\eeq
and
\beq\label{st7}
\F^{2}=\pB_{\D\gh}-\paB_{\D\gh}
\eeq

Hence $\F$ is nilpotent if the ST identites $\D\gh=0$
hold, as, for istance, for the classical action $\gh_{0}$.

If we suppose that there is a local polynomial $Q^{n}_{0}(x)$
such as
\beq\label{st11}
\Delta^{(n)}_{0}=\int \dx Q^{n}_{0}(x)~~~~~
\Delta^{(n)}_{ST}=\oF \Delta^{(n)}_{0}
\eeq
so that the breaking terms $\Delta^{(n)}_{ST}$ satisfy the consistency
\vspace{.2cm}
equation (\ref{con24}) because of nilpotence of $\oF$, then
the modified effective action:
\beq\label{st12}
\gl\stackrel{def}{=} \gh-\hbar^{n}\Delta^{(n)}_{0}
\eeq
satisfies the following equations:
\bea\label{st13} \ber{l}
\vspace{.2cm}
\pD_{\gl}\gl=\left[\meta(\pB_{\gl}+\paB_{\gl})+\aR\right]
\left( \gh-h^{n}\Delta^{(n)}_{0}\right)=
\Delta^{(n)}_{ST}-\hbar^{n}\pF_{\gh_{0}}\Delta^{(n)}_{0}+
O(\hbar^{n+1})=O(\hbar^{n+1})
\eer \eea
\ie
\beq\label{st14}
\pD_{\gl}\gl=O(\hbar^{n+1})
\eeq
and consequently solves our renormalization problem of ST identities. Obviuosly
we have to show that our hypothesis (\ref{st11}) is correct,
and this is our aim in the following.

We now enter in  a more detail analysis of breaking terms $\Delta_{ST}$,
and from the beginnings we can split them into two parts respectively dependent
and independent on s parameter:
\beq\label{rs5}
\Delta^{(n)}_{ST}=\int \dx P^{(n)}_{ST}(x) +
\sum_{k=1}^{2}(s-1)^{k}\int \dx R^{(n)}_{ST,k}(x).
\eeq

Here the local polynomials $R^{(n)}_{ST,k}(x)$, $P^{(n)}_{ST}(x)$ are
both independent on s parameter.
The polynomial $P^{(n)}_{ST}(x)$
has IR and UV degree equal to 5, whereas the $R^{(n)}_{ST,k}(x)$ terms
have $d_{R^{(n)}_{ST,k}}=r_{R^{(n)}_{ST,k}}=(5-k)$.
Since we are only interest in the s=1 theory, we have not to determine
the explicit form of $R^{(n)}_{ST,k}(x)$, but we drive our attention on
the first term of right hand side of eq. (\ref{rs5}).

One sees that, due to the dimensions and charges of the fields,
the ST breaking $\Delta^{(n)}_{ST}$ can be break up into the following
two terms:
\beq\label{rs14}
\Delta^{(n)}_{ST}=\int \dx \left(\Om^{a}_{\mu}P^{(n)\mu}_{1a}+
P^{(n)}_{2}\right)
\eeq
where $P^{(n)\mu}_{1a}$, $P^{(n)}_{2}$
are both functions of fields
$\{ A^{a}_{\mu},V^{a}_{\mu},\hg^{a}_{\mu},\om^{a},\zeta^{a} \}$
and by their derivatives.

Recalling that the external field $\Om^{a}_{\mu}(x)$ has  $\q$ = 1
and dimension 2, we can easily rewrite
the first term of r.h.s. of eq. (\ref{rs5}) in the form:
\beq\label{rs15}
\Om^{a}_{\mu}P^{(n)\mu}_{1a}=\Lambda_{abc}\Om^{a}_{\mu}\hg^{b\mu}\om^{c}+
\Om^{a}_{\mu}\tP^{\mu}_{1a}(A,V)
\eeq
where the field dependence of $\tP^{a}_{1\mu}$ is pointed out
and $\Lambda^{abc}$ is an invariant tensor under SU(N) group
transformations. The first term of (\ref{rs15}) is easily
compensated by introducing the counter term:
\beq\label{rs131}
\gh_{\Om\hg\om}\equiv \int \dx \Lambda_{abc}\hg^{a}_{\mu}\om^{b}V_{a}^{\mu}
\eeq
and observing that the following equation holds:
\beq\label{rs132}
\int \dx \Lambda_{abc}\Om^{a}_{\mu}\hg^{b\mu}\om^{c}=\aR\gh_{\Om\hg\om}.
\eeq

The expressions (\ref{rs14}) and (\ref{rs15})
allow us to reduce the consistency equation
(\ref{con24}) to a system of functional equations,
in fact  making explicit $\oF$ and $\Delta^{(n)}_{ST}$,
we get:
\bea\label{rs16} \ber{l}
\vspace{.3cm}
{\cal F}_{\gh_{0}}\Delta^{(n)}_{ST}=2\oDD\Delta^{(n)}_{ST}+
\aR\Delta^{(n)}_{ST}=\\
{}~~~~~~~=2\oDD \dms{\int} \dy \left(\Om^{a}_{\mu}P^{(n)\mu}_{1a}+
P^{(n)}_{2}\right) +\dms{\int} \dx
\left[ \Om^{a}_{\mu}\dms{\fdf{V}{a}{\mu}} \right]
\dms{\int} \dx \left(\Om^{a}_{\mu}P^{(n)\mu}_{1a}+
P^{(n)}_{2}\right).
\eer \eea
Then collecting the terms with increasing powers of
$\Om^{a}_{\mu}$,
we find:
\bea\label{rs17} \left\{ \ber{l}
\oDD \dms{\int} \dx P^{(n)}_{2}=0 \\ \vspace{.2cm}
2\oDD \dms{\int} \dx \left(\Om^{a}_{\mu}P^{(n)\mu}_{1a}\right)+
\aR \dms{\int} \dx P^{(n)}_{2}=0 \\
\dms{\int} \dx \left[ \Om^{a}_{\mu}\dms{\fdf{V}{a}{\mu}} \right]
\dms{\int} \dy \Om^{a}_{\mu}P^{(n)\mu}_{1a} =0
\eer \right. \eea
and by (\ref{rs15}) the last equation becomes:
\beq\label{rs18}
\dms{\int} \dx \Om^{a}_{\mu}(x)\dms{\fdf{V}{a}{\mu}}
\dms{\int} \dy \Om^{a}_{\mu}\widetilde{P}^{(n)\mu}_{1a} =0
\eeq

We postpone the study of this equation until the next section, because
we now  want to study the implications of system (\ref{rs17}) for the
final result.

We suppose to have been able to show that all $\Om^{a}_{\mu}$ dependent
terms of the ST breakings are compensable by a set of
$\Om^{a}_{\mu}$-independent counterterms $\{\gh_{i}\}$
\beq\label{rs23}
\int \dx \left(\Om^{a}_{\mu}P^{(n)\mu}_{1a}\right)
=\aR \sum_{i}\gh_{i}
\eeq
then modifying the effective action by
\beq\label{rs24}
\gh \rightarrow \gl\equiv \gh-\hbar^{n}\sum_{i}\gh_{i}
\eeq
we obtain the following result:
\bea\label{rs25} \ber{l}
\pD_{\gl}\gl=(\pDD_{\gl}+\aR)(\gh-\hbar^{n}\dms{\sum_{i}}\gh_{i})=
\hbar^{n}\left(\int \dx P^{(n)}_{2}-\hat{\Delta}^{(n)}_{ST}\right)+
O(\hbar^{n+1})
\eer \eea
where $\hat{\Delta}^{(n)}_{ST}\equiv 2\sum_{i}\pDD_{\gh_{i}}\gh_{0}$.
We can summarize the previous equations into the following:
\beq\label{rs270}
\pD_{\gl}\gl=\hbar^{n}\wt{\Delta}^{(n)}_{ST}+O(\hbar^{n+1})
\eeq
where the term $\wt{\Delta}^{(n)}_{ST}$ collects all $\Om$-independent
contributes coming directly from ST breakings and from
$\pDD_{\gh_{i}}\gh_{0}$. Thus we deduce that
\beq\label{rs271}
\frac{\delta \wt{\Delta}^{(n)}_{ST}}{\delta \Om^{a}_{\mu}} =0.
\eeq

Let us study the equation:
\beq\label{rs27}
\frac{\delta \gh}{\delta \Om^{a}_{\mu}} = \Delta_{\Om}\gh
\eeq
which describes the dependence of 1PI on $\Om^{a}_{\mu}$ field, and
by the means of QAP, it take in account the breakings from
tree level equation:
\beq\label{rs28}
\frac{\delta \gh_{0}}{\delta \Om^{a}_{\mu}} =0
\eeq

The breaking term $\Delta^{(n)}_{\Om}(x)$
at $\hbar^{n}$ order is a local polynomial with  $\q$= -1,
it is a vector under Lorentz transformations, a vector under global
gauge transformations and is
characterised by the Zimmerman's degrees of subtraction:
$\del_{\Delta_{\Om}}\leq 2,~~~\rho_{\Delta_{\Om}}\geq 2,~~~
\deg_{s-1}\Delta_{\Om}\geq 2$.

One immediately sees that $\Delta^{(n)}_{\Om}$ does not contain powers of
$(s-1)$ because of algebraic structure it
belongs to; moreover
it is apparent that the only possible monomial is:
\beq\label{rs30}
\Delta^{(n)a}_{\Om\mu} = \Lambda^{ab}_{\mu\nu}\hg^{\nu}_{b}
\eeq
that is a linear term in the field $\hg^{a}_{\mu}(x)$.
This fact removes the necessity of subtraction of divergences for the
operator in the r.h.s. of (\ref{rs27}) and then
this breaking term is definitively compensated by:
\beq\label{rs31}
\gh_{Eff}\rightarrow \gh_{Eff}-\int \dx
\Om^{a}_{\mu}\Lambda^{ab}_{\mu\nu}\hg^{\nu}_{b}
\eeq
recovering the equation:
\beq\label{rs32}
\frac{\delta \gh}{\delta \Om^{a}_{\mu}} =0
\eeq
at every order of the perturbative expansion. This equation is also satisfied
by the functional $\gl$, because the necessary counterterms to compensate
the $\Om$ dependent breaking terms are independent from $\Om$ itself.
This is really important because it leads to the end of our demonstration.

Let us take the derivative of both sides of equation
(\ref{rs27}) with respect to $\Om^{a}_{\mu}$ and then put it to zero:
\beq\label{rs33}
\frac{\delta}{\delta \Om^{a}_{\mu}} (\pD_{\gl}\gl)\left|_{\Om=0}\right.=
\frac{\delta}{\delta \Om^{a}_{\mu}}
\left(\wt{\Delta}^{(n)}_{ST} \right)\left|_{\Om=0}\right.
\eeq
The r.h.s. is vanishing because of (\ref{rs271}) and the first one can be
written
in the form:
\beq\label{rs34}
\frac{\delta}{\delta \Om^{a}_{\mu}} (\pD_{\gl}\gl)\left|_{\Om=0}\right.=
\frac{\delta}{\delta \Om^{a}_{\mu}}
(\pDD_{\gl}\gl+\aR\gl)\left|_{\Om=0}\right.
\eeq
that is:
\beq\label{rs35}
\frac{\delta}{\delta \Om^{a}_{\mu}} (\pD_{\gl}\gl)\left|_{\Om=0}\right.=
\frac{\delta}{\delta \Om^{a}_{\mu}} (\pDD_{\gl}\gl)\left|_{\Om=0}\right.+
\left[\dms{\fd{\gl}{V}{a}{\mu}} \right]\left|_{\Om=0}\right.=0
\eeq

Because of the eq. (\ref{rs32}) and observing that the operator
$\pDD_{\gl}$ commutes  with
the functional derivative $\frac{\delta}{\delta \Om^{a}_{\mu}}$, we have
\beq\label{rs36}
\left(\dms{\fd{\gl}{V}{a}{\mu}}\right) = 0
\eeq
for every value of the field $\Om^{a}_{\mu}$
and for every order of perturbation expansion.
This allows us to deduce that
the ST Identities are reduced to the following:
\beq\label{rs361}
\pDD_{\gl}\gl=\hbar^{n}\wt{\Delta}^{(n)}_{ST}+O(\hbar^{n+1})
\eeq
for the background field independent functional $\gl$, i.e. we
recover the conventional ST identities:
\beq\label{rs37}
\dms{\int} \dx \left[\fd{\gl}{A}{a}{\mu} \fd{\gl}{\hg}{\mu}{a} +
\fdu{\gl}{\om}{a} \fdd{\gl}{\zeta}{a} \right]=
\hbar^{n}\wt{\Delta}^{(n)}_{ST}+O(\hbar^{n+1})
\eeq

We remand  to the literature for the
study of breaking terms of eq. (\ref{rs37}). We have now to confirm our
hypotheses for the breakings of the equation (\ref{rs23}).

\subsection{Solution of consistency equation }

The open problem is to find a solution of the consistency equation for
the breakings (\ref{rs18})
\beq\label{sl1}
\dms{\int} \dx \Om^{a}_{\mu}(x)\dms{\fdf{V}{a}{\mu}}
\dms{\int} \dy \Om^{a}_{\mu}\wt{P}^{(n)\mu}_{1a}(y)=0
\eeq
in the space of local functional
$(\Om^{a}_{\mu}\wt{P}^{(n)\mu}_{1a})(y)$
linear in the external field $\Om^{a}_{\mu}(x)$,  polynomial
into the field $A^{a}_{\mu}(x)$ e $V^{a}_{\mu}(x)$ and in their derivative.
On this space we shall find that the solution can be put in the form:
\beq\label{sl2}
\dms{\int} \dy \Om^{a}_{\mu}\widetilde{P}^{(n)\mu}_{1a} =\aR \sum_{i} \gh_{i}
\eeq
It belongs to the image of $\aR$ on integrated functionals
$\gh_{i}(A,V)=\int \dx \g_{i}(x)$ neutral under $\q$ charge and
with dimension 4.

We rewrite the $\aR$ operator in a more synthetic way :
\beq\label{sl3}
\aR=\int \dx \Om^{i}(x)\frac{\delta}{\delta \phi^{i}(x)}
\eeq
and notice that, by the commutation properties of
fields $\{\Om^{i}(x)\}$ e $\{\phi^{i}(x)\}$, $\aR$ is nilpotent.

Changing a little the notation we want to give some account of
the structure of the linear space ${\cal S}$ of local functional
$\g(x)$\cite{algeb}, polynomial in  $\Om^{i}$, $\phi^{i}$
and their derivatives\footnote{
In this section we will use the notation $D_{\mu(n)}$ for
the collective derivative
$D_{\mu(n)}\Om^{i}(x)\equiv \de_{\mu_{1}}\dots\de_{\mu_{n}}\Om^{i}(x)$
and $D^{(0)}\Om^{i}(x)\equiv \Om^{i}(x)$. By the locality of
the elements of the space ${\cal S}$ these variables are functional
independent.}
$D_{\mu(n)}\Om^{i}(x)$,$D_{\mu(n)}\phi^{i}(x)$.

On ${\cal S}$ $\aR$ reduces to the local operator
\beq\label{sl4}
\aR=\sum_{n}D_{\mu(n)}\Om^{i}(x)\frac{\de}{\de D_{\mu(n)}\phi^{i}(x)}
\eeq
Since $\aR$ is nilpotent the image $Im(\aR)$ is a subspace
of its kernel
\beq\label{sl41}
{\rm Im}(\aR)\subseteq  \ker(\aR)
\eeq
then our problem  is to find out the possible solutions $\gh(x)$ of
\beq\label{sl42}
\aR\g(x)=0
\eeq
that do not belong to ${\rm Im}(\aR)$ or, in other terms,
the representatives of the cohomology classes $H(\aR)$:
\beq\label{sl43}
H(\aR)\equiv\frac{\ker(\aR)}{{\rm Im}(\aR)}
\eeq

Our space of functionals ${\cal S}$ is an Hilbert
space\footnote{More accounts in the appendix B},
so that it is possible to define  the
adjoint of $\aR$ as:
\beq\label{sl5}
\aR^{\dagger}= \sum_{n}D_{\mu(n)}\phi^{i}(x)\frac{\de}
{\de D_{\mu(n)}\Om^{i}(x)}
\eeq
and the correspondent selfadjoint Laplace-Beltrami operator:
\beq\label{sl6}
\Delta\equiv \{\aR,\aR^{\dagger}\}=
\sum_{n}\left(D_{\mu(n)}\Om^{i}(x)\frac{\de}{\de D_{\mu(n)}\Om^{i}(x)}+
D_{\mu(n)}\phi^{i}(x)\frac{\de}{\de D_{\mu(n)}\phi^{i}(x)} \right)
\eeq
where kernel ($\ker \Delta$) is the subspace of harmonic functionals:
\beq\label{sl7}
\Delta \g(x)=0
\eeq
Thanks to the Hilbert structure of ${\cal S}$ for every cohomology
class $[\g(x)]\in H(\aR)$ there is only one harmonic representative.
Then the cohomolgy class of the operator $\aR$ and of its adjoint
$\aR^{\dagger}$ is identified with the space
$\ker \Delta$, or more specifically we have:
$\ker(\aR)\cap\ker(\aR^{\dagger})=\ker\Delta$.

Form the expression of the $\Delta$ we find that the space
$\ker \Delta$ is nothing else but the space
$\hat{{\cal S}}$  of local functionals
independent on field variables $D_{\mu(n)}\Om^{i}(x)$ and
$D_{\mu(n)}\phi^{i}(x)$.  Thus every solution $\g(x)$ of
(\ref{sl42}) is given by:
\beq\label{sl8}
\g (x)=\aR\g'(x)+\gh(x)~~~~~\g'(x)\in {\cal S}~~~\gh(x)\in \hat{{\cal S}}
\eeq

We now may come back to the space of integrated functionals for which
the consistency equation is
\beq\label{sl9}
\aR \int \dx \g(x)=0
\eeq
We recall that the integration is extended to whole Euclidean space and
the conventional hypothesis of smoothness for the functionals of
${\cal S}$ are assumed. Besides the Hilbert space structure we
introduce the p-forms  $\om_{p}^{q}(x)$ with $\q$ = q and
contract the Lorentz indices
with 1-forms $dx^{\mu}$. In particular we are interested
in the solution  of (\ref{sl9}) on the 4-forms $\om_{4}^{0}(x)$ subspace.

The equation admits two types of solutions:  the cocycles and
the d-module cocycles
\beq\label{sl10}
\aR \om_{4}^{0}(x)=d \om_{3}^{1}(x)
\eeq
where d is the conventional external derivative on p-forms spaces.
Since the nilpotent operator $\aR$ commutes
with d, acting by the left on both sides of (\ref{sl10}) we obtain:
\beq\label{sl11}
(\aR)^{2} \om_{4}^{0}(x)=0=d(\aR \om_{3}^{1}(x))
\eeq
which implies that the 3-form $\om_{3}^{1}(x)$ is a d-cocycles.
By the Poincar\'e theorem we find:
\beq\label{sl12}
\aR \om_{3}^{1}(x)=d \om_{2}^{2}(x)
\eeq
i.e. the $\aR \om_{3}^{1}(x)$ form is a d-coboundary.
Iterating the process we find the following ladder of descendent equations:
\bea\label{sl13} \ber{l}
\oop{.3cm}
\aR \om_{4}^{0}(x)=d \om_{3}^{1}(x) \\
\oop{.3cm}
\aR \om_{3}^{1}(x)=d \om_{2}^{2}(x) \\
\oop{.3cm}
\aR \om_{2}^{2}(x)=d \om_{1}^{3}(x) \\
\oop{.3cm}
\aR \om_{1}^{3}(x)=d \om_{0}^{4}(x) \\
\aR \om_{0}^{4}(x)=0
\eer \eea
By the previous result (\ref{sl8}), the last equation of
(\ref{sl13}) gives:
\beq\label{sl14}
\om_{0}^{4}(x)=\aR \chi_{0}^{3}(x)+c_{0}(x)
\eeq

The 0-form $c_{0}$ is independent from variables
$D_{\mu(n)}\Om^{i}(x)$,
$D_{\mu(n)}\phi^{i}(x)$
and, since we search for local solutions,  it is a constant.
Acting with external derivative on
(\ref{sl14}) we have:
\beq\label{sl15}
d\om_{0}^{4}(x)=d\aR \chi_{0}^{3}(x)+dc_{0}=\aR(d\chi_{0}^{3})
\eeq
and then, from (\ref{sl13}), we arrive at:
\beq\label{sl16}
\aR(\om_{1}^{3}-d\chi_{0}^{3})=0.
\eeq
Again, by (\ref{sl8}), we can iterate the result:
\bea\label{sl17} \ber{l}
\oop{.3cm}
\om_{1}^{3}=d\chi_{0}^{3}+\aR\chi_{1}^{2}+c_{1} \\
\oop{.3cm}
\om_{2}^{2}=d\chi_{1}^{2}+\aR\chi_{2}^{1}+c_{2} \\
\oop{.3cm}
\om_{3}^{1}=d\chi_{2}^{1}+\aR\chi_{3}^{0}+c_{3} \\
\om_{4}^{0}=d\chi_{3}^{0}+\aR\chi_{4}^{-1}+c_{4}
\eer \eea
where the forms $c_{i}$ have constant coefficients
and then they can be written as $c_{i}=d\hat{c}_{i-1}$.
Therefore, from that sequence, we obtain:
\beq\label{sl18}
\om_{4}^{0}=\aR\chi_{4}^{-1}+c_{4}+d\chi_{3}^{0}.
\eeq
The integrated functional
\beq\label{sl19}
\int \om_{4}^{0}=\aR\int \chi_{4}^{-1}+\int c_{4}
\eeq
is the solution of consistency equation (\ref{sl9}).

The second term on the r.h.s. of the previous equation belongs to the
cohomology class of $\aR$, which we identified with the class of harmonic
functionals
independent on fields
$D_{\mu(n)}\Om^{i}(x)$, $D_{\mu(n)}\phi^{i}(x)$.
Since we are interested in the solution in the space of forms
$\om_{4}^{0}$ linear dependent on
$\Om^{a}_{\mu}$, the term $c_{4}$ necessarily vanishes.
The solution of (\ref{sl19}) allows us to express the $\Om^{a}_{\mu}(x)$-
dependent breakings of ST identities, which solve the consistency
equation  (\ref{sl1}), as \aR-variations of integrated functionals.
The analysis of the previous section shows how employ this result to
cancel up to the $\hbar^{n}$-order every  $\Om^{a}_{\mu}$-dependent
breaking of
$\Delta^{(n)}_{ST}$.
Notice also that we have not used the WT identities of background gauge
invariance. The next section is devoted to the study of these important
identities.

\subsection{Renormalization of Ward-Takahashi Identities}

We have left this set of equations as the last to be renormalized because,
as we will see, we will derive benefit from the conclusions of
the previous sections.
The necessary compensation of breakings of ST identities does not
spoil further the WT identities since we considered these latter
as broken by all
possible sorts of breaking terms. Hence we can replace the
1PI functional by the modified one $\gl$, which satisfies eq. (\ref{rs36}),
in our WT identities:
\beq\label{wt1}
W(\lambda)\gl=\hbar^{n}\Delta^{(n)}_{W}(\lambda)+O(\hbar^{n+1})
\eeq

Again we suppose the renormalizability up to the $\hbar^{n}$-order.

We have to recall the consistency equations to which the breakings
$\Delta^{(n)}_{W}(\lambda)$ must satisfy:
\beq \label{wt2}
W(\lambda)\Delta^{(n)}_{ST}-\oF \Delta^{(n)}_{W}(\lambda)=0
\eeq
and
\beq\label{wt2b}
W(\lambda)\Delta^{(n)}_{W}(\bet)-W(\bet)\Delta^{(n)}_{W}(\lambda)=
\Delta^{(n)}_{W}(\lambda\wedge\bet).
\eeq
In the first one we may also replace the ST breakings $\Delta^{(n)}_{ST}$
with the $\Om$-independent terms $\wt{\Delta}^{(n)}_{ST}$:
\beq\label{wt3}
W(\lambda)\wt{\Delta}^{(n)}_{ST}-\oF \Delta^{(n)}_{W}(\lambda)=0 \\
\eeq
It is easy to show the following commutation properties of $W(\lambda)$
and $\oF$ operators with the functional derivative with respect the
$\Om$ field:
\beq\label{wt4}
\left[ W(\lambda), \dms{\fdf{\Om}{a}{\mu}} \right]=
\left(\lambda \wedge \frac{\delta}{\delta \Om } \right)^{a}_{\mu}
{}~~~~~~\left\{ \oF, \dms{\fdf{\Om}{a}{\mu}} \right\}=
\dms{\fdf{V}{a}{\mu}}
\eeq
by which we obtain:
\bea\ber{l} \label{wt5}
\left[ W(\lambda), \dms{\fdf{\Om}{a}{\mu}} \right]
\wt{\Delta}^{(n)}_{ST}-
W(\lambda) \left( \dms{\fdf{\Om}{a}{\mu}}
\wt{\Delta}^{(n)}_{ST} \right)+
\oF \left( \dms{\fdf{\Om}{a}{\mu}} \Delta^{(n)}_{W}(\lambda) \right)
-\dms{\fdf{V}{a}{\mu}} \Delta^{(n)}_{W}(\lambda) =0
\eer \eea

Due to the results of the previous sections that our breaking terms
$\wt{\Delta}^{(n)}_{ST}$ are independent from $\Om$ we have that
the two terms of l.h.s. of equation (\ref{wt5}) vanish and then we can replace
the eq. (\ref{wt5}) with the easier one:
\beq\label{wt6}
\oF \left( \dms{\fdf{\Om}{a}{\mu}} \Delta^{(n)}_{W}(\lambda) \right)
-\dms{\fdf{V}{a}{\mu}} \Delta^{(n)}_{W}(\lambda) =0
\eeq

If we success to prove in another way that the
breakings $\Delta^{(n)}_{W}(\lambda)$ are $\Om$-independent we
may conclude:
\beq\label{wt7}
\left(\dms{\fdf{V}{a}{\mu}}\right) \Delta^{(n)}_{W}(\lambda) =0.
\eeq

We have now to make explicit the structure of $\Delta^{(n)}_{W}(\lambda)$.
In the first we split the external field $(\hg,\zeta,\Om )$ part from the
independent one:
\beq\label{wt8}
\Delta^{(n)}_{W}(\lambda)=\dms{\int} \dx
\left\{ a_{1} \tr [\lambda\hg_{\mu}\Delta^{\hg}_{\mu}]+
a_{2} \tr [\lambda\Om_{\mu}\Delta^{\Om}_{\mu}]+
a_{3} \tr [\lambda\zeta\Delta^{\zeta}]+
a_{4} \tr [\lambda\Delta(A,V)] \right\}
\eeq

Here the trace is taken over matrices of adjoint representation
so that every term is invariant under global gauge transformations.
The field $\lambda$ is a test function
carrying an index in the same representation of fields. The local
polynomials $\Delta^{\Om}_{\mu}, \Delta(A,V), \Delta^{\hg}_{\mu}$ and
$\Delta^{\zeta}$ are of increasing powers of $\q$ charge from $-1$ and
all of dimension 2 except of external field independent term
$\Delta(A,V)$ of dimension 4.

One immediately sees that the breakings $\Delta^{\Om}_{\mu}$ and
$\Delta^{\zeta}$ are of the form:
\beq\label{wt9}
(\Delta^{\zeta})^{ab}=\meta \Lambda^{ab[cd]}\om_{c}\om_{d}~~~~~
(\Delta^{\Om})^{ab}_{\mu}=\Lambda^{abc}\hg^{c}
\eeq
but these terms do not satisfy the Wess-Zumino consistency equation
(\ref{wt2b}) and then $a_{3}=a_{2}=0$.
For the $\Delta^{\hg}_{\mu}$ one finds the possible monomial decomposition:
\beq\label{wt10}
\tr [\lambda\hg_{\mu}\Delta^{\hg}_{\mu}]=
\Lambda^{abc}_{1}\lambda^{a}\de^{\mu}(\hg^{b}_{\mu}\om^{c})+
\Lambda^{abc}_{2}\lambda^{a}\hg^{b}_{\mu}\de^{\mu}\om^{c}+
\Lambda^{abcd}_{3}\lambda^{a}\hg^{b}_{\mu}V^{c}_{\mu}\om^{d}+
\Lambda^{abcd}_{4}\lambda^{a}\hg^{b}_{\mu}A^{c}_{\mu}\om^{d}
\eeq

It is direct to show that only the first term satisfy the
Wess-Zumino equation (\re{wt2b}) then we have to seek, if exists,
its counter term, but we have:
\beq\label{wt11}
W(\beta)\dms{\int} \dx tr[\om\hg_{\mu}V_{\mu}]=-\dms{\int} \dx
tr[\beta \de_{\mu}(\hg_ {\mu}\om)]
\eeq

Thus introducing this counter term into
the effective action up to the $\hbar^{n}$-order,
we have only to consider the external field independent terms
$\Delta(A,V)$.
One has to note also that this counter term
spoils again the ST identities, but, by the theorem proved
before, we have always the possibility to adjust finely the coefficients of
counter terms in order to cancel the $\Om^{a}_{\mu}$-dependent breakings
both for ST and for WT identities.
By the previous result (\re{wt7}), we have that $\Delta(A,V)$ is independent
from the background field $V^{a}_{\mu}$ and then the only possible anomalies
is the conventional ABJ (\cite{abj}).

{\large \bf Acknowledgements.}

I am deeply indebted to Prof. C.M.Becchi for a
constant help and supervision during all steps of the draft of this paper.
Particular thanks to Prof. C.Destri for a careful reading of the manuscript,
and to Prof\`\,s L.Girardello and M.Martellini for useful discussions.

\vskip 1 true cm
\noindent
\begin{center}
{\large \bf Appendix }
\end{center}
\label{coho}
In this appendix we want to describe shortly the Hilbert structure
of the ${\cal S}$ of local polynomial functionals used in the
previous sections, then we want to give just a little account about
the Hodge decomposition theorem on this space.

We have been forced to introduce the base of the space ${\cal S}$
defined by independent variables $D_{\mu(n)}\Om^{i}(x)$ and
$D_{\mu(n)}\phi^{i}$ with opposite commutative character, on the
same ground we introduce the adjoint base:
$D_{\mu(n)}\Om^{i\dagger}(x)$, $D_{\mu(n)}\phi^{i\dagger}$
with the relations:
\bea\label{app1} \ber{l}
\left\{D_{\mu(m)}\Om^{i}(x),D_{\nu(n)}\Om^{j\dagger}(x)\right\}=
\del_{m,n}\del_{ij}\del^{\mu(1)\dots\mu(m)}_{\nu(1)\dots\nu(n)} \\
\left[D_{\mu(m)}\phi^{i}(x),D_{\nu(n)}\phi^{j\dagger}(x)\right]=
\del_{m,n}\del_{ij}\del^{\mu(1)\dots\mu(m)}_{\nu(1)\dots\nu(n)}
\eer \eea
where
$\del^{\mu(1)\dots\mu(m)}_{\nu(1)\dots\nu(n)}$ is a notation for
symmetrised Kronecker symbol. From this definition it is easy
to show that $\aR$ is the adjoint operator, and to show that
Laplace-Beltrami operator
\beq\label{app2}
\Delta\equiv\aR\aR^{\dagger}+\aR^{\dagger}\aR
\eeq
is selfadjoint.

{\it Decomposition theorem}. The vector space
${\cal S}$ could be spitted into direct sum of orthogonal subspace:
\beq\label{app3}
{\cal S}=Im(\aR)\oplus Im(\aR^{\dagger}) \oplus
(\ker(\aR)\cap\ker(\aR^{\dagger}))
\eeq
{\it Proof}. From the decomposition:
\beq\label{app4}
{\cal S}=Im\Delta \oplus \ker\Delta
\eeq
every vector $\om$ can be written in the form:
\beq\label{app5}
\om=\Delta\chi+\xi=\{\aR\aR^{\dagger}+\aR^{\dagger}\aR\}\chi+\xi
\eeq
where $\Delta\xi=0$, and such as, if we set
$\chi'=\aR^{\dagger}\chi$,
$\chi''=\aR\chi$, we have
\beq\label{app6}
\om=\Delta\chi+\xi=\aR\chi'+\aR^{\dagger}\chi''+\xi
\eeq
and, by
$0=(\xi,\Delta\xi)=(\aR\xi,\aR\xi)+(\aR^{\dagger}\xi,\aR^{\dagger}\xi)$
we have immediately:
\beq\label{app7}
\aR\xi=\aR^{\dagger}=0
\eeq
i.e. our thesis.

\eject
\end{document}